\PassOptionsToPackage{nameinlink}{cleveref}
\documentclass[a4paper,USenglish,cleveref]{lipics-v2021}
\hideLIPIcs  % arXiv preprint: no LIPIcs logo, DOI, event or article number
\nolinenumbers
\usepackage{amsmath,amssymb,amsthm}
\usepackage[T1]{fontenc}
\usepackage{graphicx}
\usepackage{booktabs}
\usepackage{xcolor}
\usepackage[ruled,vlined,linesnumbered]{algorithm2e}
\allowdisplaybreaks[2]
\makeatletter
\@fleqnfalse
\makeatother

\newboolean{showcomments}
\setboolean{showcomments}{false}
\ifthenelse{\boolean{showcomments}}
{ \newcommand{\mynote}[3]{
     \fbox{\sffamily\tiny#1}
       {\small$\blacktriangleright$\textcolor{#3}{{ #2}}}}}
       {\newcommand{\mynote}[3]{}}

\newif\ifshowtodo
\showtodofalse
\ifshowtodo
  \newcommand{\todo}[1]{{\color{red}TODO: #1}}
\else
  \newcommand{\todo}[1]{}
\fi

\newcommand{\IntervalCount}{150{,}461}
\newcommand{\ReferenceFee}{0.500369}
\newcommand{\ReferenceCost}{0.000197}
\newcommand{\CostCorrelation}{0.979}
\newcommand{\CostCV}{3.79}
\newcommand{\CostPeak}{417}
\newcommand{\MeanIntervalSeconds}{120.87}
\newcommand{\ExecutionShare}{98.6}
\newcommand{\ArbitrumGain}{0.111766}
\newcommand{\ArbitrumEta}{0.012582912}

\newcommand{\PRMSE}{0.0529}

\newcommand{\PMove}{2.14}

\newcommand{\PIRMSE}{0.41}

\newcommand{\PIMove}{7.63}

\newcommand{\LQRRMSE}{0.0249}

\newcommand{\LQRMove}{4.9}

\newcommand{\LQRHighRMSE}{0.00331}
\newcommand{\LQRHighMove}{36.2}
\newcommand{\LQRHighPeak}{296}
\newcommand{\ArbitrumHighRMSE}{0.00486}
\newcommand{\ArbitrumHighMove}{23.8}
\newcommand{\ArbitrumHighPeak}{215}
\newcommand{\ConstantRMSE}{4.83}

\newcommand{\FigureWidth}{1.0\linewidth}
\newcommand{\TableWidth}{.5\linewidth}
\title{Optimal Dynamic Fees for L2s: A Vault Approach}

\author{Ulysse {Pavloff}}{University of Bern, Switzerland}{ulysse.pavloff@unibe.ch}{https://orcid.org/0000-0003-4125-3306}{}
\author{Lin Oshitani}{Nethermind Research}{}{}{}

\authorrunning{U. Pavloff and L. Oshitani}
\Copyright{U. Pavloff and L. Oshitani}



\crefname{observation}{Observation}{Observations}
\Crefname{observation}{Observation}{Observations}

\begin{document}

\maketitle

% ---------------------------------------------------------------

\begin{abstract}
Layer 2 (L2) blockchains incur costs when posting data to Layer 1 (L1) and charge users fees intended to recover those costs.
We model this mechanism as a \textit{vault}: user fees increase its balance, L1 posting costs reduce it, and surplus and deficit accumulate over time.
Forecast-based fees aim to recover posting costs, but forecast errors accumulate in the vault. Returning the vault balance to a target requires feedback on the accumulated error.
Treating the vault as a discrete-time control system, we derive in closed form the optimal trade-off between vault tracking and fee smoothness in an idealized benchmark.
The optimal rule retains a fraction of the previous fee and corrects in proportion to the current vault error.
This fee-memory fraction is the single parameter tracing the entire trade-off frontier.
We compare proportional (P), proportional-integral (PI), and Arbitrum-style controllers against this frontier, both analytically and on historical Ethereum fee data.
At the fee movement implied by Arbitrum's current parameters, the optimal rule reduces the benchmark tracking error by about one third.
Fee memory improves the tracking--smoothness trade-off.
\end{abstract}

% ---------------------------------------------------------------

\section{Introduction}
\label{sec:introduction}

In early April 2026, Scroll raised its L1 posting-cost oracle multipliers six times in six days, compounding to $1{,}280\times$ the baseline. Users paid over \$50{,}000 for transactions that would have cost about \$280 under the original parameters \cite{defiant_2026_scroll}. The L1 price readings were accurate. The excess came from manually adjusted multipliers: observing costs accurately does not ensure that fees recover the intended amount.

Scroll prices the L1 posting cost from external information, such as L1 prices and expected data volume. This is a \textit{feedforward} mechanism. A \textit{feedback} rule instead uses the current difference between the vault balance and its target. With a feedback rule, this excess would have raised the balance above its target, and the fee would then have decreased.
We develop a benchmark for these fee rules with two objectives: vault tracking and fee smoothness.

We model the accounting position of an L2 (also referred to as a rollup) as a \textit{vault}. User fees increase its balance, and L1 posting costs decrease it. Recovering the posting cost is a flow problem; keeping the accumulated balance near a target is a stock problem. 

For example, suppose the target is $10$ ETH and the vault contains $8$ ETH. If posting costs $1$ ETH in the next interval and user fees collect exactly $1$ ETH, the vault still contains $8$ ETH. Recovering that interval's cost preserves the existing shortfall. To reduce it, fee revenue must temporarily exceed posting expenditure.

Only instantaneous pass-through makes each interval pay for itself, and it requires knowing the coming interval's cost and demand. With fixed demand, it passes cost fluctuations directly into fees and leaves an existing surplus or deficit unchanged. Smoothing or imperfect prediction leaves residuals, and residuals accumulate. Feedforward anticipates costs; feedback reads the resulting balance error and corrects it.

Because the vault carries each revenue--cost mismatch forward, its dynamics form a \textit{discrete integrator}. We define metrics to quantify tracking and fee smoothness.
The \textit{integral squared error} (ISE) measures the time-averaged squared balance deviation from its target, normalized by that target. The \textit{quadratic variation} (QV) measures the time-averaged squared change between consecutive fees. We also report the \textit{worst-case drawdown} (WCD), the largest shortfall (balance below target), normalized by that target. WCD measures downside exposure separately; it is not part of the objective we optimize.

Under the idealized benchmark of \cref{sec:objective-benchmark}, we derive the optimal ISE--QV trade-off using a \textit{linear-quadratic regulator} (LQR). The optimal fee rule retains a fraction of the previous fee and corrects in proportion to the current vault error. The retained fraction, which we call the \textit{fee-memory coefficient}, is the single parameter tracing the entire frontier.

With a penalty on fee changes, fee increments accumulate into the fee, and the fee accumulates into the vault error. This structure is a discrete double integrator, whose LQR solution is classical. We use it to answer three practical questions.
\begin{itemize}
  \item \textbf{Is cost recovery enough?} No. Fees set from cost information alone cannot remove an accumulated surplus or deficit; the vault balance must feed back into the fee (\cref{sec:theory-ff}).
  \item \textbf{What is the best feedback rule, and how far are deployed rules from it?} In an idealized benchmark (\cref{sec:objective-benchmark}), the optimal rule keeps part of the previous fee and corrects the vault error. The share it keeps sets how fast the vault returns to its target versus how smooth the fees are. Arbitrum Nitro's rule is a PI controller; at its fee movement, the optimal rule has about one third less tracking error (\cref{sec:theory-lqr,sec:theory-pareto}).
  \item \textbf{What survives on real prices?} On seven months of Ethereum prices, fee memory keeps its tracking advantage at smoother settings, while proportional feedback has the smaller worst-case drawdown, and a fee cap makes the reserve size decisive (\cref{sec:evaluation}).
\end{itemize}
\Cref{sec:related-work,sec:model} give the context and the model. \Cref{sec:theory} derives the optimum and analyzes P, PI, and Arbitrum-style controllers against it. \Cref{sec:evaluation} compares them on historical Ethereum prices, and \cref{sec:discussion} draws design guidance and open problems.

\section{Related Work}
\label{sec:related-work}

\subparagraph{Transaction Fee Mechanism Design.}
\label{par:tfm}

Roughgarden \cite{roughgarden_2024_tfm} formalizes incentive requirements for blockchain transaction fee mechanisms, while Chung and Shi \cite{chung_2023_foundations} establish limitations on satisfying them jointly. The dynamical perspective is closer to our problem: Leonardos et al.\ \cite{leonardos_2021_dynamical} analyze convergence and instability of EIP-1559, and Reijsbergen et al.\ \cite{reijsbergen_2021_honeymoon} study its empirical behavior. These works examine fee adjustment in relation to block utilization; we examine it in relation to a target vault balance.

\subparagraph{Optimal Control and L2 Pricing.}
\label{par:dynamic-pricing}

Crapis et al.\ \cite{crapis_2024_optimal} provide the closest methodological precedent. They combine squared utilization error with squared price changes and derive linear-quadratic policies under deterministic price sensitivity. They approximate the stochastic-sensitivity case and compare simpler fee rules. We instead regulate a balance that accumulates fee revenue minus posting costs.

Wang et al.\ \cite{wang_2025_combined} combine posting decisions, transaction queues, fee-sensitive demand, and pricing for budget balance and congestion control. Their budget-balancing rule already adjusts fees from profit or loss accumulated over a posting period. We fix cadence and demand and regulate the balance carried across periods around a vault target. This yields an explicit frontier between vault tracking and fee smoothness.

\subparagraph{L2 Fee Mechanisms.}
\label{par:rollups}

%The documented L1 fee formulas of OP Stack and Scroll use L1 prices, estimates of transaction data size, and configurable parameters \cite{opstack_2026_docs,scroll_2025_docs}. These provide examples of feedforward pricing: the formulas estimate posting costs without reading an accumulated reserve error. 
Most major rollups price the L1 cost by feedforward: they estimate it at sequencing time and pass the estimate through to the user, typically as a relayed L1 base fee and blob base fee multiplied by operator-set scalars \cite{opstack_2026_docs,base_2026_docs,scroll_2025_docs,zksync_2024_docs,starknet_2026_docs}. A vault contract may exist for accounting, as with the L1 Fee Vault predeploy
of OP Stack that collects the L1-cost component for withdrawal to a configured recipient, but its
balance is not fed back into the price.

Arbitrum Nitro is the closest deployed antecedent to our vault model \cite{arbitrum_2024_docs,nitro_2024_source}. It collects posting fees in a reimbursement pool and adjusts pricing using the surplus, defined as the pool balance minus outstanding obligations, and the change in that surplus. Feedback on an accumulated balance is therefore already deployed. We adapt this feedback structure to the common vault target and compare its stability and performance with P, PI, and the optimal benchmark.

\section{System Model}
\label{sec:model}

A \emph{controller}, or \emph{policy}, is a rule choosing the fee $F_\tau$ for each interval. Its \emph{gains} are the coefficients multiplying its inputs.

\subsection{Discrete-Time Vault Dynamics}
\label{sec:model-dynamics}

\subparagraph{Posting intervals.}
Time $\tau\in\mathbb{N}$ indexes fixed posting intervals. Fees accumulate throughout each interval, and the L1 posting cost is paid once, at the end of the interval. Payment times are fixed but amounts remain random. A single update accounts for both revenue and expenditure.

The four primary interval-level signals are:
\begin{itemize}
  \item $V_\tau\in\mathbb{R}$, the accounting balance of the vault at the start of interval $\tau$;
  \item $F_\tau\in\mathbb{R}$, the posting-cost component of the L2 fee per data unit charged during interval $\tau$;
  \item $\Lambda_\tau\in\mathbb{R}_+$, the aggregate number of charged data units during the interval;
  \item $C_\tau\in\mathbb{R}_+$, the aggregate L1 cost of posting the batch associated with the interval.
\end{itemize}
Here $F_\tau$ is the posting-cost component of the L2 fee alone. Execution, congestion and any other fee components are outside the model.

The vault balance evolves according to the \emph{vault recurrence} equation:
\begin{equation} \label{eq:vault-recurrence}
  V_{\tau+1}=V_\tau+F_\tau\Lambda_\tau-C_\tau.
\end{equation}

\subparagraph{What a negative vault means.}
$V_\tau$ is an accounting position, not necessarily a smart-contract token balance. A negative value records obligations exceeding assigned assets. The consequences of exhausting on-chain funds lie outside these accounting dynamics.

\subparagraph{Demand and fee constraints.}
The closed-form analysis assumes constant, price-independent charged volume, $\Lambda_\tau=\bar\Lambda>0$, which isolates vault regulation from demand variation and fee sensitivity. This is a modeling simplification, not an empirical claim that total L2 demand is insensitive to fees; \cref{sec:discussion-elastic} discusses the available evidence and the extension to elastic demand.

The analytical benchmark permits real-valued fees, including negative fees. The historical evaluation applies a zero fee floor and no upper cap. Clipping the LQR rule forfeits its unconstrained optimality guarantee.

\subparagraph{Reporting scope.}
The main formal analysis and historical evaluation assume immediate reporting
of L1 posting costs to L2. Reported and economic balances then coincide, and
we simply call $V_\tau$ the vault balance. Cost-reporting delay is treated in
\cref{sec:discussion-reporting} and \cref{app:delay}.

\subparagraph{Target and equilibrium coordinates.}
The protocol specifies an exogenous accounting target $V^*>0$, which we take as given rather than deriving it from a solvency requirement; when backed by dedicated funds, it is also the intended funding level. We use $V^*$ as the normalization scale; arbitrary targets, including zero, can be treated with a separate positive scale.

Let $\bar C:=\mathbb{E}[C_\tau]$ denote the mean posting cost, assumed constant in the analysis.
Under fixed aggregate demand $\bar\Lambda$, flow balance fixes the break-even fee:
\begin{equation}
  \label{eq:Fstar}
  F^*=\frac{\bar C}{\bar\Lambda}.
\end{equation}
Thus $V^*$ sets the stock target and $F^*$ the break-even flow.

Let us define the normalized vault error and fee deviation:
\begin{equation}
  \label{eq:error-coordinates}
  \epsilon_\tau:=\frac{V^*-V_\tau}{V^*},
  \qquad
  \tilde F_\tau:=F_\tau-F^*,
  \qquad
  \gamma:=\frac{\bar\Lambda}{V^*}.
\end{equation}
A balance below target gives $\epsilon_\tau>0$, a \emph{shortfall}. A surplus gives $\epsilon_\tau<0$.
The constant $\gamma$ is the fee's leverage on the vault. It translates a fee deviation into normalized revenue: $\tilde F_\tau\bar\Lambda$ is extra revenue per interval, and $\gamma\tilde F_\tau$ is that revenue as a fraction of the target.
Under immediate reporting, substituting \cref{eq:Fstar,eq:error-coordinates} into \cref{eq:vault-recurrence} gives the exact deviation dynamics:
\begin{equation}
  \label{eq:linearised}
  \epsilon_{\tau+1}
  =\epsilon_\tau-\gamma\tilde F_\tau
   +\frac{C_\tau-\bar C}{V^*}.
\end{equation}
No local linearization is used here; the only structural simplification is the inelastic-demand assumption.
Writing:
\begin{equation}
  \label{eq:integrator}
  u_\tau:=\gamma\tilde F_\tau,
  \qquad
  w_\tau:=\frac{C_\tau-\bar C}{V^*},
\end{equation}
gives the \emph{controlled integrator}:
\begin{equation}
  \label{eq:controlled-integrator}
  \epsilon_{\tau+1}=\epsilon_\tau-u_\tau+w_\tau.
\end{equation}
Each interval adds the cost disturbance ($w_\tau$) and subtracts the fee correction ($u_\tau$) from the accumulated error.
The coefficient on $\epsilon_\tau$ is the system's \emph{pole}: here it is exactly one, so the existing error does not decay on its own but is carried forward until a fee correction offsets it. \cref{eq:controlled-integrator} is exact under fixed demand.

\subsection{L1 Cost Process}
\label{sec:model-l1}

\subparagraph{Units and cost decomposition.}
At posting-interval resolution $C_\tau$ is a \emph{total} monetary cost. 
A simple decomposition is $C_\tau=G_\tau B_\tau+D_\tau P^{\mathrm{blob}}_\tau$, execution gas $G_\tau$ times its price $B_\tau$ plus blob gas $D_\tau$ times its price $P^{\mathrm{blob}}_\tau$; each product is in currency (e.g., ETH).
We model posting costs using execution and blob base fees, excluding priority fees (tips).

\subparagraph{Exogeneity and mean.}
The L1 posting cost $C_\tau$ is exogenous to the fee controller. Wang et al.~\cite{wang_2025_combined} also take L1 prices as given, but optimize the posting decisions that determine expenditure; we fix the cadence and treat the cost sequence as an external input.
Strategic posting and reimbursement incentives may invalidate this assumption (\cref{sec:discussion-open}).

The mean $\bar C=\mathbb E[C_\tau]$ is finite and constant. It fixes $F^*$ in \cref{eq:Fstar} and centers the disturbance $w_\tau=(C_\tau-\bar C)/V^*$. \Cref{sec:objective-benchmark} states the remaining stochastic assumptions; the evaluation tests the controllers under persistent fluctuations, mean shifts, and spikes.

\subsection{Performance Metrics}
\label{sec:model-metrics}

The mechanism has two primary objectives: keep the vault balance close to its target $V^*$ and avoid abrupt fee changes. We also report the largest shortfall below target.

\begin{definition}[Vault regulation and maximum shortfall]
  \label{def:vault-health}
  Over a horizon $T$, the \emph{integral squared error} is:
  \begin{equation}
    \label{eq:ise}
    \mathrm{ISE}_T:=\frac{1}{T}\sum_{\tau=1}^{T}\epsilon_\tau^2,
  \end{equation}
  and the \emph{worst-case drawdown} is:
  \begin{equation}
  \label{eq:wcd}
  \mathrm{WCD}_T:=\max\left\{0,\max_{1\le\tau\le T}\epsilon_\tau\right\}.
    \end{equation}
\end{definition}

ISE penalizes deviations above and below the target $V^*$ equally, and quadratically: doubling a deviation quadruples its contribution.
WCD measures only the largest shortfall, below the fixed target: it is zero if the vault never falls below target, equals one if its minimum balance is zero, and exceeds one if the balance becomes negative.
ISE and QV are the design metrics; WCD reports downside severity.\footnote{Adding WCD to the objective would make the objective depend on the maximum error over the whole horizon, rather than only on additive quadratic costs, so the LQR solution would no longer apply directly. We optimize ISE and QV to obtain an exact benchmark and report WCD separately to assess severe shortfalls.}

\begin{definition}[Fee smoothness]
  \label{def:fee-volatility}
  The \emph{quadratic-variation rate} of the fee path is
  \begin{equation}
    \label{eq:qv}
    \mathrm{QV}_T:=\frac{1}{T}\sum_{\tau=1}^{T}
      (F_\tau-F_{\tau-1})^2.
  \end{equation}
\end{definition}

QV measures fee movement, not dispersion around a mean. The square penalizes a few abrupt jumps more heavily than many small changes, and it is the smoothness criterion of the LQR benchmark.

All three metrics describe a single trajectory over $T$ posting intervals and require no probabilistic assumptions. \Cref{sec:theory} characterizes the long-run expectations of ISE and QV. \cref{sec:evaluation} computes the three metrics by replaying historical costs through each controller.

\subsection{Control Objective and Benchmark}
\label{sec:objective-benchmark}

The closed-form results rest on four assumptions, which we refer to by label.
\begin{description}
  \item[(A1) Fixed demand.] Charged volume is constant, $\Lambda_\tau=\bar\Lambda$.
  \item[(A2) Known mean cost.] The baseline $F^*=\bar C/\bar\Lambda$ uses the true, constant mean cost.
  \item[(A3) Unpredictable cost deviations.] The disturbance satisfies \cref{eq:disturbance-assumption} below, given the vault and fee history.
  \item[(A4) Unrestricted fees.] Fees may take any real value, including negative ones.
\end{description}
The historical evaluation violates three of them: costs are persistent and spiky (A3), their mean drifts (A2), and fees have a zero floor (A4). Demand stays fixed (A1); \cref{sec:discussion-elastic} discusses this choice.

\subparagraph{Benchmark information and disturbance assumption.}
The analytical benchmark gives the controller the current vault balance
and the past vault and fee history. Let:
\begin{equation}
  \label{eq:lqr-filtration}
  \mathcal H^{\mathrm{LQ}}_\tau
  :=\sigma\bigl(\epsilon_0,\dots,\epsilon_\tau;\;
     \tilde F_{-1},\dots,\tilde F_{\tau-1}\bigr)
\end{equation}
be this history. No contemporaneous L1 signals that could predict $C_\tau$ are supplied. 
For the closed forms, assume that the normalized disturbance $w_\tau$ of \cref{eq:integrator} is a square-integrable martingale difference: it has a finite expected square and conditional mean zero given this history. 
We also assume a constant conditional variance relative to this history:
\begin{equation}
  \label{eq:disturbance-assumption}
  \mathbb E\bigl[w_\tau\mid\mathcal H^{\mathrm{LQ}}_\tau\bigr]=0,
  \qquad
  \mathbb E\bigl[w_\tau^2\mid\mathcal H^{\mathrm{LQ}}_\tau\bigr]
  =\sigma_w^2<\infty.
\end{equation}
Thus the history provides no systematic prediction of the next cost deviation, and the uncertainty about that deviation, its variance, stays constant. 
I.i.d. zero-mean disturbances satisfy these conditions, but full independence is not required.

\subparagraph{Unconstrained quadratic benchmark.}
We ask how accurately a controller can regulate the vault for a given penalty on fee changes. The benchmark supplies the correct break-even baseline $F^*=\bar C/\bar\Lambda$ and removes fee bounds. We assume that the initial error $\epsilon_0$ and previous fee
deviation $\tilde F_{-1}$ have finite second moments. Let $\mathcal M_{\mathrm{LQ}}$ be the class of controllers whose chosen fee deviations $\tilde F_\tau$ are square-integrable and adapted to the history \cref{eq:lqr-filtration}: each fee is chosen using information available at that time. No future observations are available when a fee is chosen.

We also require finite time-averaged second moments of the vault error and fee deviation, as specified in \cref{app:proofs}. This permits fluctuations without allowing their average expected squared magnitude to diverge. The class permits negative fees; its optimum is a reference for the unconstrained problem. For a controller $\mu\in\mathcal M_{\mathrm{LQ}}$ and weights $q,r>0$, define the \emph{quadratic objective} function:
\begin{equation}
  \label{eq:lqr-cost}
  J(\mu;q,r)
  :=\limsup_{T\to\infty}\frac{1}{T}
    \mathbb E\!\sum_{\tau=0}^{T-1}
    \left[q\epsilon_\tau^2+r(F_\tau-F_{\tau-1})^2\right].
\end{equation}
With nonzero persistent noise, even a stable controller incurs cost indefinitely, so we use an average rather than the diverging undiscounted sum.\footnote{The $\limsup$ takes the upper limiting value if a policy's running average does not converge. For the stationary optimum, the ordinary limit exists.} Only the relative weight $q/r$ affects the optimizer.

\subparagraph{ISE--QV benchmark frontier.}
Within the benchmark, a Pareto frontier describes trade-offs at which neither vault tracking nor fee smoothness can improve without worsening the other. Varying $q/r$ selects which objective matters more and traces the \emph{supported} frontier: points minimizing $q\,\mathrm{ISE}+r\,\mathrm{QV}$ for positive weights.

\section{Optimality and Stability}
\label{sec:theory}

The observation below needs no stochastic assumptions. All later results assume (A1)--(A4).

\subsection{Cost Recovery and Balance Regulation}
\label{sec:theory-ff}

Whenever feasible, exact pass-through sets  $F_\tau=C_\tau/\Lambda_\tau$, so fee revenue equals the interval's posting cost. This still leaves $V_{\tau+1}=V_\tau$. 
It cannot correct an existing surplus or shortfall.
More generally, $V_T-V_0=\sum_{\tau<T}(F_\tau\Lambda_\tau-C_\tau)$: the balance records cumulative revenue minus costs.

\begin{observation}[Feedforward cannot regulate an unknown level offset]
\label{prop:feedback-necessary}
Consider two vaults with different initial balances, but exposed to identical costs, charged volumes, and external information. A fee rule using only that information produces the same fees in both runs. 
Therefore, both balances cannot converge to the same target. Feedforward alone cannot correct an initial balance offset.
\end{observation}

\begin{proof}
The rule produces identical fees in both runs. Subtracting the
vault recurrence \cref{eq:vault-recurrence}, identical revenues
and costs cancel at each interval, giving
\[
  V_\tau^{(1)}-V_\tau^{(2)}
  =V_0^{(1)}-V_0^{(2)}
  \qquad\text{for every }\tau.
\]
\end{proof}

The observation does not say that feedforward can never compensate a known cost shock; it says that flow information alone cannot regulate an independent accumulated level error. It applies to the pass-through designs, whatever their calibration.

Even with a constant break-even fee:  $\epsilon_{\tau+1}=\epsilon_\tau+w_\tau$, for fixed initial error we have:
\[
 \mathbb E[\mathrm{ISE}_T]=\epsilon_0^2+\frac{T+1}{2}\sigma_w^2.
\]
The expected ISE grows linearly when $\sigma_w^2>0$, and a nonzero constant mean error adds quadratic growth. Stable proportional feedback keeps second moments bounded. 
The implication is asymmetric: any controller that keeps the vault bounded recovers the flow on average, while pass-through shows that the converse fails. 
A deployed rule can combine both, and the benchmark asks how well the feedback part can possibly do.

\subsection{LQR-Optimal Benchmark}
\label{sec:theory-lqr}

Minimizing the quadratic objective \cref{eq:lqr-cost} subject to the integrator dynamics \cref{eq:controlled-integrator} is a linear-quadratic regulator problem with a penalty on control increments. Its exact optimum belongs to the following family.

\subparagraph{Leaky-integrator family.}
Choose a memory fraction $\pi\in(0,1)$ and a correction gain $k>0$. The fee rule is:
\begin{equation}
\label{eq:leaky-family}
F_\tau=\pi F_{\tau-1}+k\,\epsilon_\tau+(1-\pi)F^*.
\end{equation}
A shortfall contributes an upward correction $k\epsilon_\tau$, although the fee can still fall if its previous value is high. Unrolling gives
\[
 \tilde F_\tau=\pi^{\tau+1}\tilde F_{-1}
   +k\sum_{j=0}^{\tau}\pi^j\epsilon_{\tau-j}.
\]
Each past error loses a fraction $1-\pi$ of its weight per step. We call this a \emph{leaky integrator} and $\pi$ its \emph{fee-memory coefficient}. At fixed $k$, the endpoints $\pi=0$ and $\pi=1$ give proportional and pure integral action, respectively. The optimum instead ties $k$ to $\pi$.

\subparagraph{Including the previous fee in the state.}
The same vault error can call for different fee adjustments depending on the fee already being charged. To account for the penalty on fee changes, we therefore include the previous fee deviation in the state, $(\epsilon_\tau,\tilde F_{\tau-1})$, and choose the fee increment.
This gives a standard LQR problem. \cref{app:proofs} states the matrices and the Riccati equation used to solve it.

\begin{theorem}[Optimal controller for the unconstrained benchmark]
\label{thm:lqr}
Among the controllers in $\mathcal M_{\mathrm{LQ}}$, the quadratic objective \cref{eq:lqr-cost} is minimized by the following member of the leaky-integrator family \cref{eq:leaky-family}:
\begin{equation}
  \label{eq:lqr-policy}
  F_\tau=\pi F_{\tau-1}+k\,\epsilon_\tau+(1-\pi)F^*,
  \qquad
  k=\sqrt{\frac{q\pi}{r}}.
\end{equation}
For given weights $q,r>0$, the optimal memory coefficient $\pi$ is the unique solution in $(0,1)$ of
\begin{equation}
  \label{eq:lqr-quartic}
  (1-\pi)^4=\frac{q}{r}\gamma^2\pi(1+\pi)^2.
\end{equation}
Its minimum average cost is
$J^*=r\sigma_w^2(1-\pi)^3/\bigl(\gamma^2\pi(1+\pi)\bigr)$.
The gain depends on the dynamics and relative objective weights, but not on the disturbance variance: changing $\sigma_w^2$ scales the achieved cost without changing the optimal rule.
\end{theorem}

\begin{proof}[Proof idea]
The Riccati equation selects the gain that balances future vault error against fee movement. 
Completing the square shows that no admissible controller has a lower average cost.
Solving for the gains gives \cref{eq:lqr-quartic}, with a unique root because its sides vary in opposite directions on $(0,1)$. The full argument is in \cref{app:proofs}.
\end{proof}

\subparagraph{The memory coefficient is the design knob.}
Choosing $\pi$ directly determines the optimal gain: combining \cref{eq:lqr-policy,eq:lqr-quartic} gives
\begin{equation}
  \label{eq:lqr-gain-pi}
  k=\frac{(1-\pi)^2}{\gamma(1+\pi)}.
\end{equation}
The closed-loop poles, which determine how initial disturbances decay, have modulus $\sqrt\pi$. When fee movement is cheap ($q/r\to\infty$), $\pi\to0$ and the rule becomes $u_\tau=\epsilon_\tau$: it removes the existing error in one step, leaving only the new disturbance $\epsilon_{\tau+1}=w_\tau$. This is \emph{deadbeat} control and attains the minimum ISE. When fee movement is expensive ($q/r\to0$), $\pi\to1$ and the optimal gain tends to zero; fees move less and errors persist longer.

\subparagraph{Damping of the optimal loop.}
Taking the fee memory into account gives the fee inertia. After a shock, the fee can still be above $F^*$ when the vault reaches its target. This overshoot that arrives before settling is at most about $4.3\%$ of the initial error (computed numerically for fee memory $\pi\in(0,1)$). Conversely, $P$ with $\kappa_p\le1$ never overshoots.

\subparagraph{Fee memory comes from penalizing fee changes.}
Imagine that instead of penalizing fee changes ($(F_{\tau}-F_{\tau-1})^2$), the objective penalized the fee deviation from the break-even fee: $\tilde F_\tau^2=(F_\tau-F^*)^2$. 
In this version, the previous fee would no longer impact future costs. The state would simply be $\epsilon_\tau$, and the optimal rule would be proportional, $u_\tau=\kappa\epsilon_\tau$, with $\kappa\in(0,1)$ given by a scalar Riccati equation. 
Thus, fee memory is the signature of a penalty on fee changes, not on fee levels. Penalizing both would give a family of rules that contains P and the leaky integrator as special cases (\cref{sec:discussion-open}).

\subparagraph{Supported ISE--QV benchmark frontier.}
Varying $\pi\in(0,1)$ gives the \emph{optimal tracking and fee-movement curves}:
\begin{align}
  \mathrm{ISE}(\pi)
  &=\sigma_w^2\frac{(1+\pi)(1+4\pi+\pi^2)}
    {(1-\pi)(1+6\pi+\pi^2)},
  \label{eq:lqr-ise}\\
  \mathrm{QV}(\pi)
  &=\frac{2\sigma_w^2(1-\pi)^3}
    {\gamma^2(1+\pi)(1+6\pi+\pi^2)}.
  \label{eq:lqr-qv}
\end{align}
For $\sigma_w^2>0$, as $\pi$ rises from zero to one, ISE rises from its floor $\sigma_w^2$ to infinity and QV falls from $2\sigma_w^2/\gamma^2$ to zero; together they satisfy $q\,\mathrm{ISE}+r\,\mathrm{QV}=J^*$. This curve is the supported frontier of \cref{sec:objective-benchmark}, and restricted families evaluated under the same assumptions cannot lie below it.

\subparagraph{Implementing a zero fee floor.}\label{par:floor}
The historical evaluation applies a nonnegative fee with no upper cap and stores that applied fee for the next update.
The leaky rule has no separate accumulator in which an uncharged correction can build up.
The floor changes the dynamics, so the unconstrained optimality guarantee no longer applies.
It also makes regulation asymmetric: since $V_{\tau+1}\ge V_\tau-C_\tau$, a surplus can only drain at the rate of posting costs, whatever the controller.

\subparagraph{Constant mean cost mismatch.}\label{par:mean-mismatch}
The leaky integrator does not fully reject a constant error in the calibrated mean. If the true mean cost exceeds the baseline used in $F^*$ by $\Delta$, the disturbance mean is $\Delta/V^*$ and the steady-state error is
\[
  \epsilon^*
  =\frac{1-\pi}{k}\frac{\Delta}{\gamma V^*}
  =\frac{1+\pi}{1-\pi}\frac{\Delta}{V^*}.
\]
The offset is smallest at the aggressive end, $\Delta/V^*$ as $\pi\to0$, and grows without bound as $\pi\to1$: the controller is least robust to persistent mean mismatch exactly where it is smoothest.
Integral action, introduced next, can maintain a correction after the error has returned to zero.
It can therefore help when mean costs shift, although a constant-mismatch calculation does not predict a trajectory with changing costs and a fee floor.

\subsection{Feedback Controllers: P, PI, and the Arbitrum-Style Rule}
\label{sec:ctrl-feedback}

The optimum assumes an accurately calibrated mean cost and unrestricted fees.
We compare the optimal rule with a proportional (P) controller, a proportional-integral (PI) controller, and an adaptation of Arbitrum Nitro's fee rule with fixed parameters (the closest deployed posting fee mechanism to our vault model). 
We first compare these rules within the analytical benchmark; \cref{sec:evaluation} then examines their performance under changing cost conditions and a nonnegative-fee constraint.

We write $K_p$ and $K_i$ for the proportional and integral fee gains.
Their corresponding normalized loop gains are $\kappa_p:=\gamma K_p$ and $\kappa_i:=\gamma K_i$.
The factor $\gamma$ converts a fee correction into its effect on the normalized vault error.

\subparagraph{P: only correct the current vault error.}
The \emph{proportional fee rule} is
\begin{equation}
  \label{eq:p-controller}
  F_\tau=F^*+K_p\epsilon_\tau.
\end{equation}
Substitution into the normalized control definition gives
$u_\tau=\gamma(F_\tau-F^*)=\kappa_p\epsilon_\tau$.
The baseline supplies the revenue needed even when the vault is on target. If the true mean cost exceeds the calibrated mean by $bV^*$, the stable loop leaves a mean error $\epsilon^*=b/\kappa_p$: P needs a continuing shortfall to maintain the extra fee.

\subparagraph{PI: retain the correction for a persistent mismatch.}
The \emph{proportional-integral fee rule} adds an accumulator,
\begin{align}
  I_{\mathrm{acc},\tau}
    &=I_{\mathrm{acc},\tau-1}+\epsilon_\tau,\notag\\
  F_\tau
    &=F^*+K_p\epsilon_\tau+K_i I_{\mathrm{acc},\tau}.
  \label{eq:pi-controller}
\end{align}
PI retains the full accumulated contribution of past errors, whereas the leaky rule gives progressively smaller weights to older errors. A persistent shortfall therefore keeps raising its stored correction until the fee covers the true mean cost. In a stable, unsaturated loop, the mean fixed point has $\epsilon^*=0$ and $I_{\mathrm{acc}}^*=b/\kappa_i$. The correction remains even when the current error is zero. Fee memory improves the benchmark tracking--smoothness trade-off; integral action instead rejects persistent baseline error.

Subtracting consecutive updates without clipping cancels the
constant baseline $F^*$ and gives the \emph{incremental form}:
\[
 F_\tau-F_{\tau-1}
 =K_p(\epsilon_\tau-\epsilon_{\tau-1})+K_i\epsilon_\tau.
\]
This form updates the previous fee without an explicit baseline
or separate accumulator: the previous fee carries the accumulated
correction. The Arbitrum-style rule below has this form with a
specific relationship between its proportional and integral gains.

The historical evaluation uses the accumulator form of PI with a
zero fee floor. 
When the fee is clipped at zero, the PI accumulator can continue integrating negative balance errors even though the requested negative fee cannot be applied. 
This stored correction can keep the fee at zero even after the vault falls below target, delaying recovery.
This effect is called \emph{integrator windup}.
To limit this windup, we reject an update of the accumulator when both the resulting requested fee and the current
error are negative, then recompute the fee and apply the zero
floor. Positive errors are still accumulated to \emph{unwind} the
negative correction.
This rule can also freeze the accumulator away from the floor. After a large surplus, one update can lower the requested fee by $|K_i\epsilon_\tau|$, more than the current fee. Every update would then make the requested fee negative and is rejected, thus the applied fee stays positive. By doing so, the surplus is not removed: PI's vault stays in surplus for weeks after 10 October (\cref{fig:spike}). The historical PI results describe PI with this anti-windup rule, not PI in general.

\subparagraph{Arbitrum Nitro: update the fee from surplus.}
Nitro adjusts its posting price from the reimbursement pool's surplus, defined as its balance minus outstanding obligations \cite{arbitrum_2024_docs,nitro_2024_source}. We adapt this structure to the common target, writing $S_\tau=V_\tau-V^*$. Its parameter $\mathrm{EquilUnits}$ is a volume over which to correct surplus, and $\mathrm{Inertia}=\nu$ sets the update factor. All volumes below are expressed in the same charged unit as $\bar\Lambda$; Nitro's pricing units must first be converted when that unit is a byte.

At a report event, the implementation updates its accounting surplus and then sets the price used subsequently. In the immediate-reporting interval model, the boundary balance $S_\tau$ therefore determines $F_\tau$, the fee for the upcoming interval. With constant update volume $\bar U=\bar\Lambda$, the unconstrained \emph{surplus fee update} is
\begin{equation}
 \label{eq:arb-update}
 \begin{aligned}
 F_\tau-F_{\tau-1}
 &=c\left(-\frac{S_\tau}{\mathrm{EquilUnits}}
         -\frac{S_\tau-S_{\tau-1}}{\bar U}\right),\\
 c&=\frac{\bar U}{\mathrm{EquilUnits}/\nu+\bar U}.
 \end{aligned}
\end{equation}
The first term corrects the accumulated surplus; the second responds to its change per charged unit. Inertia scales their combined fee increment, rather than supplying a separate derivative term. Summing the increments gives a PI rule (\cref{app:arbitrum}) with
\[
 \eta:=\frac{\bar\Lambda}{\mathrm{EquilUnits}},\qquad
 \kappa_p^{\mathrm A}=c=\frac{\nu\eta}{1+\nu\eta},\qquad
 \kappa_i^{\mathrm A}=\eta c.
\]
Thus the equilibration volume fixes the ratio of integral to proportional action; inertia varies their magnitude. Like incremental PI, this rule can retain the required fee level without an explicit $F^*$. Nitro clamps its posting price at zero; this is separate from the minimum execution fee charged on L2.

\subparagraph{Stability of the three controllers.}
A controller can amplify the imbalance it is meant to correct if its gains are too large. Stability analysis identifies the gain ranges in which the effect of an initial imbalance fades and second moments remain bounded under the disturbance assumptions. These conditions make the long-run ISE and QV comparisons meaningful; they do not guarantee that the vault can never be depleted. For P, the error recursion is $\epsilon_{\tau+1}=(1-\kappa_p)\epsilon_\tau+w_\tau$: each correction leaves a fraction $1-\kappa_p$ of the previous error. Its magnitude is below one exactly for $\kappa_p\in(0,2)$. PI is stable exactly when
\begin{equation}
  \label{eq:pi-stable}
  \kappa_p\in(0,2),\qquad \kappa_i>0,\qquad 2\kappa_p+\kappa_i<4.
\end{equation}
Stronger proportional action leaves less room for integral action. At $2\kappa_p+\kappa_i=4$, a closed-loop root reaches $-1$ and an alternating component stops decaying. The Arbitrum adaptation with immediate feedback obeys the same condition, with its tied gains: $c(2+\eta)<4$.

\subparagraph{Benchmark tracking and fee movement.}
For P, a small gain changes fees gently but lets disturbances persist in the vault. Increasing the gain toward $\kappa_p=1$ reduces tracking error at the cost of greater fee movement. Under (A1)--(A4), these quantities are
\[
 \mathrm{ISE}=\frac{\sigma_w^2}{1-(1-\kappa_p)^2},
 \qquad
 \mathrm{QV}=\frac{2\sigma_w^2K_p^2}{2-\kappa_p}.
\]
At $\kappa_p=1$, P removes the existing error in one interval and reaches the minimum ISE, $\sigma_w^2$. Gains between one and two remain stable but overcorrect: both ISE and QV increase. We therefore compare P only on $\kappa_p\in(0,1]$, where improving tracking requires greater fee movement.

The PI expressions are given in \cref{eq:pi-ise,eq:pi-qv}. The Arbitrum adaptation with immediate feedback is a PI controller with $\kappa_p=c$ and $\kappa_i=\eta c$, so its metrics follow by substituting those gains into the same expressions.

As $\kappa_i\to0$, the accumulator contributes progressively less to the fee, and PI's limiting metrics tend to P's. At zero integral gain, the rule is exactly P for the same baseline. Both metrics increase strictly with $\kappa_i$ at fixed $\kappa_p$, so each PI point is dominated by P at that proportional gain. The benchmark supplies the correct mean: the persistent error that integral action would remove is absent. Its extra cost buys robustness to departures from that assumption.

\subsection{Benchmark Comparison}
\label{sec:theory-pareto}

\begin{figure}[tbp]
  \centering
  \includegraphics[width=\FigureWidth]{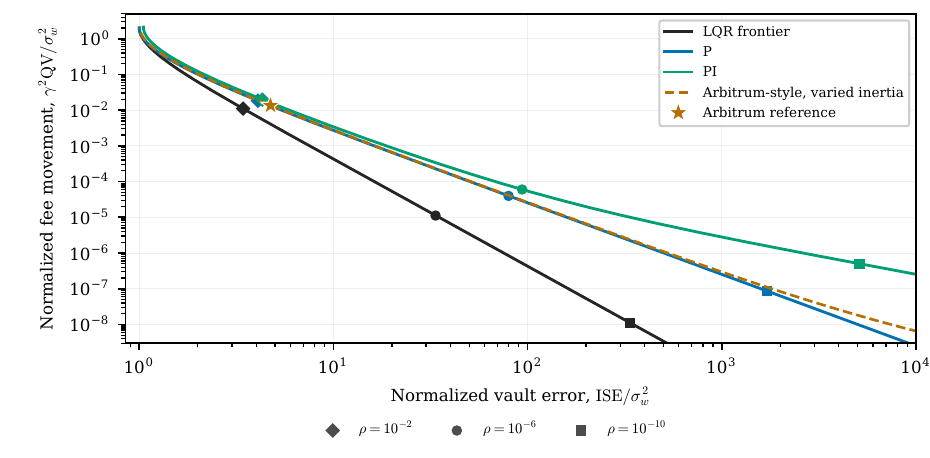}
  \caption{Unconstrained benchmark with immediate feedback. Markers identify the relative weight on vault
  tracking versus fee smoothness: $\rho\in\{10^{-2},10^{-6},10^{-10}\}$ on P, fixed-ratio PI and LQR. PI fixes $\kappa_i/\kappa_p=0.1$. The dashed Arbitrum-style curve varies inertia continuously at fixed $\eta\simeq0.0126$; only the star retains the reference inertia $10$. Points farther down and to the left have smaller fee movement and vault error, respectively. These are analytical moments under the benchmark assumptions, not historical outcomes.}
  \label{fig:frontier}
\end{figure}

\Cref{fig:frontier} places the rules in normalized coordinates: $X=\mathrm{ISE}/\sigma_w^2$ for vault error and $Y=\gamma^2\mathrm{QV}/\sigma_w^2$ for fee movement, for $\sigma_w^2>0$. We retain the weights $q,r$ of \cref{eq:lqr-cost} and define the dimensionless calibration ratio:
\[
  \rho:=\gamma^2\frac{q}{r}\;\;,\qquad
  q\,\mathrm{ISE}+r\,\mathrm{QV}
  =\frac{r\sigma_w^2}{\gamma^2}(\rho X+Y).
\]
The positive scale factor leaves the minimizing controller unchanged. The experiments minimize $\rho X+Y$ at $\rho\in\{10^{-2},10^{-6},10^{-10}\}$, corresponding to physical weight ratios $q/r=\rho/\gamma^2$. ISE is an outcome of calibration rather than a prescribed matching level.
For PI, we fix the ratio $\kappa_i/\kappa_p=0.1$ and optimize $\kappa_p$.\footnote{Fixing the ratio helps retain nonzero integral action. Under the benchmark, increasing $\kappa_i$ at fixed $\kappa_p$ increases both ISE
and QV; so optimizing both gains would drive $\kappa_i$
toward zero, recovering P.}
The dashed Arbitrum-style curve varies inertia continuously at a fixed gain ratio; it is a mathematical relaxation rather than a set of deployed configurations. The star is the single fixed reference with the workload-derived ratio $\eta=\ArbitrumEta$ and fixed inertia, using the same parameters as the historical evaluation (\cref{sec:eval-setup}). It is not retuned to any objective weight.

P reaches the frontier only in the limiting deadbeat case, at $(1,2)$, where the optimum needs no fee memory. Elsewhere, retaining the previous fee reduces movement at matched accuracy. At the smooth end, error grows like $\mathrm{QV}^{-1/3}$ along the frontier but like $\mathrm{QV}^{-1/2}$ for P.
PI approaches P's curve as the integral gain tends to zero. Its gap combines the restriction to the PI family with the cost of positive integral gain under correct mean calibration.

\subparagraph{Where the Arbitrum reference stands.}
As can be seen in \cref{fig:frontier}, the Arbitrum reference closely follows the P controller. This stems from the fact that its integral term is small: $\eta\approx0.013$. P with the same fee movement has $X=4.70$, against $4.74$ for the reference. Theoretically, the reference can be improved by reducing $X$, $Y$, or both. But every rule on the frontier is a leaky integrator with $\pi>0$; P reaches it only at the deadbeat point. Any frontier rule that improves on the reference therefore uses fee memory. At the same fee movement, it reduces tracking error by $32\%$; at the same tracking, it needs $70\%$ less $QV$.

\section{Evaluation}
\label{sec:evaluation}

\subsection{Setup}
\label{sec:eval-setup}

\subparagraph{Observed prices and modeled workload.}
The historical evaluation uses Ethereum prices between Pectra (7 May 2025) and Fusaka (3 December 2025), one blob-fee regime. It contains \IntervalCount\ globally aligned intervals of ten L1 blocks, averaging $\MeanIntervalSeconds$ seconds. For each interval, prices at its final block determine the modeled posting cost: $150{,}000$ execution gas plus three blobs of $131{,}072$ blob-gas units each. Charged volume is fixed at $393{,}216$ modeled byte units, an allocation convention rather than a claim about usable blob capacity or Nitro compression. These base-fee-only expenditures omit execution priority fees; they are not recorded Arbitrum spending.

Mean cost is $\bar C=\ReferenceCost$ ETH, its coefficient of variation is \CostCV, and the largest cost is \CostPeak\ times the mean. Execution gas contributes \ExecutionShare\% of mean cost. Adjacent interval costs have correlation \CostCorrelation, so (A3) fails: the last cost strongly predicts the next. Spikes also violate its constant variance, and a drifting mean violates (A2). The historical evaluation is therefore a stress test of the benchmark rules, not a validation of the benchmark.

\subparagraph{Initialization and reference parameters.}
All runs begin at $V_0=V^*=10$ ETH and $F_{-1}=F_{\rm ref}=\bar C/\bar\Lambda=\ReferenceFee$ gwei per modeled byte, with no further controller memory. Fees have a zero floor and no upper cap. P, PI, and LQR use this constant baseline; incremental Arbitrum has no baseline input. The full-window mean is a shared retrospective calibration, not an estimate available at deployment. The target is illustrative, not a derived funding requirement. 

The Arbitrum-style reference uses equilibration volume $E=500{,}000{,}000$ pricing units and inertia $\nu=10$, giving $\eta=\ArbitrumEta$ and $c=\ArbitrumGain$ after conversion at 16 pricing units per modeled byte. These parameters were verified on 10 September 2026; they define a fixed reference, not a reconstruction of the deployed 2025 rule. The source mapping and parameter provenance are in \cref{app:arbitrum}.

All rules process the completed interval's revenue and cost before choosing the next fee. Real Nitro report arrivals and allocated volumes are not reconstructed; \cref{app:delay} analyzes fixed reporting delay separately.

\subparagraph{Objective weights and descriptive sweep.}
For each of P, PI with $\kappa_i=0.1\kappa_p$, and LQR, three marked settings minimize that family's theoretical objective at $\rho\in\{10^{-2},10^{-6},10^{-10}\}$. The weights are illustrative, and the PI results concern this fixed gain ratio only.
The main table reports all three weights; the middle weight, $10^{-6}$, supplies the stress traces. \Cref{tab:weights} lists the gains and theoretical coordinates. 
\Cref{fig:tradeoff} shows a dense sweep of 601 weights per family (\cref{app:sweep}).

\subparagraph{Reported metrics.}
Vault RMSE is $V^*\sqrt{\mathrm{ISE}_T}$ in ETH; maximum shortfall is $V^*\mathrm{WCD}_T$ in ETH. RMS fee change is $100\sqrt{\mathrm{QV}_T}/F_{\rm ref}$ percent per posting interval. Peak fee is $\max_{0\le\tau<T}F_\tau/F_{\rm ref}$: the largest charged fee, measured separately from its changes. The reference is fixed throughout the historical evaluation. \Cref{app:sweep} specifies the boundary indexing, including the terminal fee decision.

\subsection{Results}
\label{sec:eval-results}

% GENERATED by experiments.paper_revision; edit the generator, not this file.
\begin{table}[tbp]
\centering\small
\setlength{\tabcolsep}{3.2pt}
\caption{Historical outcomes with a shared zero fee floor and no upper cap. Each weight calibrates P, fixed-ratio PI and LQR to their own theoretical objective. Arbitrum retains its fixed parameters and has no tuning weight. RMSE and maximum shortfall are in ETH; RMS fee change is a percentage of $F_{\rm ref}$; peak fee is a multiple of $F_{\rm ref}$. Smaller values mean less error, shortfall, fee movement or peak exposure, respectively; the columns do not define an overall winner.}
\label{tab:main}
\resizebox{\TableWidth}{!}{%
\begin{tabular}{@{}llrrrr@{}}
\toprule
Controller & $\rho$
& \shortstack{RMSE\\(ETH)}
& \shortstack{Shortfall\\(ETH)}
& \shortstack{RMS change\\(\% of $F_{\rm ref}$)}
& \shortstack{Peak fee\\($F/F_{\rm ref}$)} \\
\midrule
P & $10^{-2}$ & 0.00501 & 0.323 & 25.3 & 216 \\
PI & $10^{-2}$ & 0.00701 & 0.273 & 34.6 & 302 \\
LQR & $10^{-2}$ & 0.00331 & 0.284 & 36.2 & 296 \\
\midrule
P & $10^{-6}$ & 0.0529 & 1.15 & 2.14 & 37.9 \\
PI & $10^{-6}$ & 0.41 & 0.879 & 7.63 & 142 \\
LQR & $10^{-6}$ & 0.0249 & 0.957 & 4.9 & 93 \\
\midrule
P & $10^{-10}$ & 0.445 & 1.31 & 0.109 & 2.95 \\
PI & $10^{-10}$ & 0.472 & 1.65 & 0.653 & 28.5 \\
LQR & $10^{-10}$ & 0.126 & 1.61 & 0.293 & 13.5 \\
\midrule
Arbitrum ref. & -- & 0.00486 & 0.338 & 23.8 & 215 \\
Constant & -- & 4.83 & 7.92 & 0 & 1 \\
\bottomrule
\end{tabular}%
}
\end{table}

\begin{figure}[tbp]
  \centering
  \includegraphics[width=\FigureWidth]{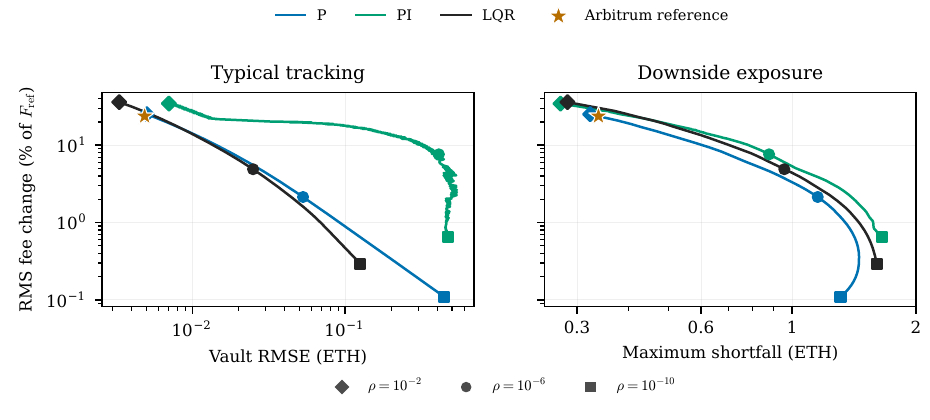}
  \caption{Dense historical sweep with a zero fee floor and no upper cap. Each curve connects 601 settings of the weight $\rho$ (\cref{app:sweep}), with objective weights marked; Arbitrum is one fixed reference. Bends reflect the historical evaluation, including changes in which event determines maximum shortfall. These curves are not empirical optimal frontiers.}
  \label{fig:tradeoff}
\end{figure}

\subparagraph{Tracking and fee movement must be read together.}
At $\rho=10^{-2}$, LQR has RMSE $\LQRHighRMSE$ ETH, versus $\ArbitrumHighRMSE$ for the fixed Arbitrum reference, but RMS fee change is $\LQRHighMove\%$ versus $\ArbitrumHighMove\%$ of $F_{\rm ref}$. Although this LQR point has both smaller error and smaller fee movement in the analytical benchmark, its historical fee movements are larger: the theoretical ordering is not a guarantee in this historical evaluation (\cref{sec:eval-setup}).

At the middle weight, $10^{-6}$, LQR improves both RMSE ($\LQRRMSE$ versus $\PIRMSE$ ETH) and fee movement ($\LQRMove\%$ versus $\PIMove\%$) over PI, whose error mostly comes from the surplus its anti-windup rule holds after 10 October. P moves fees less ($\PMove\%$) but has RMSE $\PRMSE$ ETH. The fixed Arbitrum setting has smaller RMSE but greater fee movement than LQR. The constant fee has RMSE $\ConstantRMSE$ ETH despite retrospective break-even calibration.

\subparagraph{Downside and peak fees add information.}
RMSE averages deviations on both sides of the target. It shows neither the worst deficit nor the fees paid to avoid it. Both can change the ranking. Across the whole sweep, P has a lower maximum shortfall than LQR at equal fee movement (\cref{fig:tradeoff}, right), although LQR tracks at least as well (left). A likely reason is that P raises the fee as soon as the error appears, whereas fee memory spreads LQR's response over several intervals. A low RMSE can also be bought with fee spikes: with no upper cap, the high-weight LQR and fixed Arbitrum runs reach fees of $\LQRHighPeak$ and $\ArbitrumHighPeak$ times $F_{\rm ref}$.

\begin{figure}[tbp]
  \centering
  \includegraphics[width=\FigureWidth]{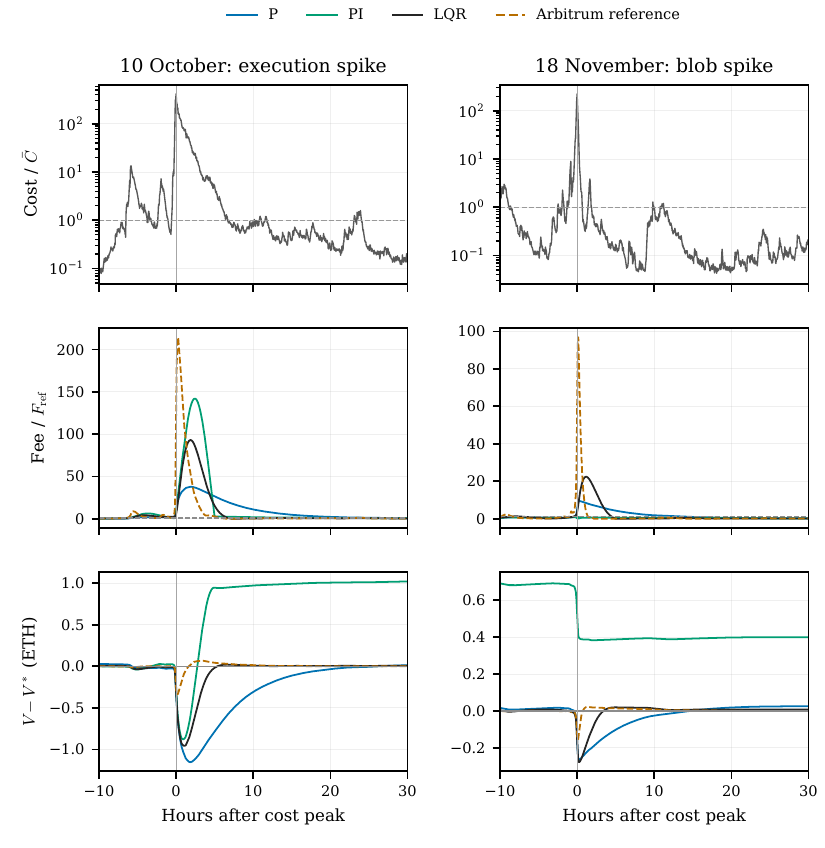}
  \caption{Stress episodes with a zero fee floor and no upper cap, at $\rho=10^{-6}$ for P, PI and LQR and the fixed Arbitrum reference. October is the left column and November the right; from top to bottom, rows show modeled posting cost, fee and post-posting vault balance relative to target. The fee baseline is dashed. Each display spans 10 hours before to 30 hours after the cost peak, without resetting the run.}
  \label{fig:spike}
\end{figure}

\subparagraph{Two stress events.}
\Cref{fig:spike} pairs the execution-cost spike of 10 October with the blob-cost spike of 18 November. Both are slices of the continuous main run, with no reset or gain change. Each column shows the cost shock, fee response and resulting balance from 10 hours before to 30 hours after the cost peak. The differing fee and balance paths explain why average error, maximum shortfall and fee movement should be reported separately. PI's lasting surplus after 10 October comes from its anti-windup rule (\cref{sec:ctrl-feedback}).

\subsection{Fee Cap}
\label{sec:eval-robustness}

% GENERATED by experiments.paper_revision; edit the generator, not this file.
\begin{table}[tbp]
\centering\small
\setlength{\tabcolsep}{3.2pt}
\caption{Sensitivity to the shared upper cap $10F_{\rm ref}$, retaining the zero floor. Gains, costs and initialization match Table~\ref{tab:main}; units are the same. Cap and Zero are percentages of charged intervals at those bounds. PI can reject an accumulator update before the applied fee reaches a bound, so these frequencies do not measure all effects of clamping.}
\label{tab:capped}
\resizebox{\TableWidth}{!}{%
\begin{tabular}{@{}llrrrrrr@{}}
\toprule
Controller & $\rho$ & RMSE & Shortfall & RMS change & Peak fee & Cap & Zero \\
\midrule
P & $10^{-2}$ & 0.0734 & 1.44 & 6.6 & 10 & 1.95 & 0 \\
PI & $10^{-2}$ & 0.0733 & 1.43 & 7.87 & 10 & 2.04 & 0 \\
LQR & $10^{-2}$ & 0.0733 & 1.44 & 8.14 & 10 & 2.11 & 0.0226 \\
\addlinespace[2pt]
P & $10^{-6}$ & 0.0836 & 1.47 & 1.53 & 10 & 0.699 & 0 \\
PI & $10^{-6}$ & 0.0744 & 1.42 & 3.14 & 10 & 1.04 & 0 \\
LQR & $10^{-6}$ & 0.0762 & 1.46 & 2.17 & 10 & 1.3 & 0.293 \\
\addlinespace[2pt]
P & $10^{-10}$ & 0.445 & 1.31 & 0.109 & 2.95 & 0 & 0 \\
PI & $10^{-10}$ & 0.18 & 1.65 & 0.485 & 10 & 0.0286 & 0 \\
LQR & $10^{-10}$ & 0.129 & 1.61 & 0.272 & 10 & 0.278 & 0.116 \\
\addlinespace[2pt]
Arbitrum ref. & -- & 0.0733 & 1.43 & 6.02 & 10 & 1.45 & 0 \\
Constant & -- & 4.83 & 7.92 & 0 & 1 & 0 & 0 \\
\bottomrule
\end{tabular}%
}
\end{table}

\subparagraph{A fee cap dominates the controller.}
\Cref{tab:capped} adds a shared upper cap of $10F_{\rm ref}$ to the zero floor. When $\rho\ge10^{-6}$, every rule then has RMSE between 0.073 and 0.084 ETH and maximum shortfall between 1.42 and 1.47 ETH, the fixed Arbitrum reference included. The cap binds during the October spike, and no rule can recover faster than the cap allows. The reserve size and the cap, not the feedback law, then determine the downside. And when $\rho=10^{-10}$, the cap rarely binds and the uncapped ordering remains.

\section{Discussion}
\label{sec:discussion}

\subparagraph{Design guidance.}
\label{sec:discussion-guidance}
Two points follow from the analysis and the historical evaluation.
\begin{enumerate}
  \item \emph{Vault feedback is necessary.} Flow information cannot remove a level error (\cref{prop:feedback-necessary}). A constant fee drifts far from target even when calibrated in hindsight.
  \item \emph{There is a trade-off between smooth fee changes and smooth fee levels.} Retaining part of the previous fee is optimal when fee changes are penalized, while a proportional rule is optimal when deviations from the break-even fee are (\cref{sec:theory-lqr}). No rule is optimal for both.
\end{enumerate}

\subparagraph{Calibration and target.}
\label{sec:discussion-calibration}
The benchmark assumes a known, constant mean cost (A2). If it is wrong, P and the leaky rule settle at a balance offset, which grows with fee memory (\cref{par:mean-mismatch}); stable, unsaturated PI removes it, and Arbitrum-style updates need no explicit mean. A deployed rule could also estimate the baseline from past posting costs; designing this estimator jointly with the controller lies beyond our fixed-mean analysis. Finally, our analysis takes the vault target $V^*$ as given. A natural extension is to choose the target and controller jointly, balancing the cost of committed funds, fee smoothness, and depletion risk.

\subparagraph{Cost reporting on L2.}
\label{sec:discussion-reporting}
An L1 posting cost is paid when the batch is posted, but the L2 learns it only when the cost report arrives, $\delta$ intervals later. This $\delta$ depends on the mechanism, from 0 for a centralized sequencer that sees the cost immediately, to arbitrarily long.
Meanwhile, the L2's ledger already counts the fees but not the delayed cost: the \emph{reported} balance looks healthier than the \emph{economic} balance by the costs not yet reported.
Regulating the reported balance keeps the same optimal gains and ISE--QV trade-off, but leaves the economic balance about $\delta\bar C$ below target on average. Raising the reported target by $\delta\bar C$ restores the economic balance to its target on average, with the same gains (\cref{app:delay}). The delay still costs some tracking accuracy, because the unreported costs are uncertain. Variable delays and batched reporting remain open.

\subparagraph{Limitations and open problems.}
\label{sec:discussion-open}
The evaluation covers one L1 price trace and one blob-fee regime, with a fixed posting cadence and payload, and with gains selected from the theoretical formulas. It does not optimize posting times or solve the control problem with a fee floor.

The main open problem is predictable cost. Adjacent costs have correlation \CostCorrelation, so the last observed L1 price carries most of the information about the next cost, yet the benchmark withholds it. Modeling the cost deviation as an observed autoregressive state keeps the problem linear-quadratic; the optimal fee would then combine the leaky feedback of \cref{thm:lqr} with a feedforward term in the observed cost. Another extension is to penalize both fee changes and deviations from the break-even fee, which would give a family of rules containing P and the leaky integrator as special cases.

\subparagraph{Demand and strategic posting.}
\label{sec:discussion-elastic}
Demand that responds to price is another possible extension, not modeled here. Evidence of weak sensitivity of gas use conditional on transaction count \cite{anchuri_2026_elasticity} does not establish that total charged volume is insensitive to the posting fee component. If higher fees reduce demand, they may weaken the revenue response assumed here. Strategic posting can also make costs endogenous when the reimbursed party influences expenditure. Both questions require behavioral models beyond retuning the present controller.

\section{Conclusion}
\label{sec:conclusion}

We modeled the accounting position of an L2 as a vault. The vault is charged by fee revenue, and discharged by posting costs. We studied how setting the fees based on the vault could make fee revenue match posting costs over time, without overcharging users or running a deficit, while limiting fee movement. 
Under (A1)--(A4), the optimal fee rule retains part of the previous fee and corrects the current vault error. A single setting, the share of the previous fee it retains, covers every optimal compromise between tracking and smoothness.
Compared with these optimal rules, the deployed Arbitrum rule behaves like a proportional controller. In the benchmark, at the same fee movement, the optimal rule has about a third less tracking error.

We evaluated these rules on seven months of Ethereum prices. Fee memory keeps its tracking advantage at smoother settings, while proportional feedback has the smaller worst-case drawdown. Introducing an upper fee cap makes the reserve size decisive.

Further research should look into rules taking the L1 price into account, so that feedforward and feedback are designed together.

\subparagraph{Use of AI tools.}
Large language models were used throughout the research process as interactive research tools, in particular for brainstorming, discussing possible approaches, reviewing the literature, and assisting with code used to generate figures and verify proofs. The authors remain responsible for the correctness and interpretation of all results presented in this paper.

\clearpage
\bibliographystyle{plainurl}
\bibliography{references}

\appendix
\crefalias{section}{appendix}

\section{Proof of Theorem~\ref{thm:lqr}}
\label{app:proofs}

\subparagraph{State and Riccati equation.}
To price fee changes, augment the vault error with the previous fee deviation:
$z_\tau=(\epsilon_\tau,\tilde F_{\tau-1})^\top$.
With control $v_\tau=\Delta\tilde F_\tau$, the deviation dynamics become
\[
 z_{\tau+1}=Az_\tau+Bv_\tau+B_ww_\tau,\qquad
 A=\begin{pmatrix}1&-\gamma\\0&1\end{pmatrix},\quad
 B=\begin{pmatrix}-\gamma\\1\end{pmatrix},\quad
 B_w=\begin{pmatrix}1\\0\end{pmatrix}.
\]
The top row records the effect of the carried fee and its increment on the vault; the bottom row stores the new fee. The stage cost is $z_\tau^\top Qz_\tau+rv_\tau^2$, where $Q=\operatorname{diag}(q,0)$.

The pair $(A,B)$ is controllable since $\det[B\;AB]=\gamma\ne0$, and $(Q^{1/2},A)$ is observable for $q>0$. Thus the discrete algebraic Riccati equation has a unique stabilizing solution $P$ \cite{bertsekas_2017_dp}:
\[
 P=Q+A^\top PA-A^\top PB(r+B^\top PB)^{-1}B^\top PA,
 \qquad K^*=(r+B^\top PB)^{-1}B^\top PA.
\]
The quadratic form $z^\top Pz$ records future excess cost above the long-run average, and the equation makes that accounting consistent with one optimally controlled step. Write $K^*=(k_1^*,k_2^*)$. The calculation below gives $k_1^*=-k$, $k_2^*=1-\pi$, and
$P_{11}=r(1-\pi)^3/\bigl(\gamma^2\pi(1+\pi)\bigr)$, so $J^*=P_{11}\sigma_w^2$.

\begin{proof}[Proof of \cref{thm:lqr}]
Write $B^\top PA=(m_1,m_2)$ and $x=B^\top PB$, so $K^*=(m_1,m_2)/(r+x)$.
Expansion gives $m_2=x$, hence $k_2^*=x/(r+x)$; the retained fraction is
therefore $\pi=1-k_2^*=r/(r+x)$, that is $x=r(1-\pi)/\pi$.
The $(1,1)$ entry of the Riccati equation reads
$q=m_1^2/(r+x)$, so $k_1^*=-\sqrt{q/(r+x)}=-\sqrt{q\pi/r}=-k$ and
$m_1=-\sqrt{qr/\pi}$, the signs fixed by stability. The $(2,2)$ and $(1,2)$
entries give $p_{22}=x-x^2/(r+x)=r(1-\pi)$ and $p_{12}=m_1\pi=-\sqrt{qr\pi}$.
Substituting these into $m_1=-\gamma p_{11}+p_{12}$ and into
$x=\gamma^2p_{11}-2\gamma p_{12}+p_{22}$ and eliminating $p_{11}$ yields
\cref{eq:lqr-quartic}. Its left-hand side decreases from one to zero and its right-hand side increases from zero on $(0,1)$, so the root is unique.
The policy $\Delta\tilde F_\tau=-K^*z_\tau$ reads $\tilde F_\tau-\tilde F_{\tau-1}=k\epsilon_\tau-(1-\pi)\tilde F_{\tau-1}$, which is \cref{eq:lqr-policy}.
It remains to prove optimality over the whole class, not only over linear rules.
The moment condition defining $\mathcal M_{\mathrm{LQ}}$ is \[  \limsup_{T\to\infty}\frac1T\sum_{\tau=0}^{T-1}  \mathbb E\!\left[\epsilon_\tau^2+                    \gamma^2\tilde F_{\tau-1}^2\right]<\infty. \] The fixed factor $\gamma^2$ expresses both terms in normalized vault units. Let $h(z):=z^\top Pz$ and write $v_\tau:=\Delta\tilde F_\tau$ for the fee increment chosen by an arbitrary controller in $\mathcal M_{\mathrm{LQ}}$.
Completing the square in the Riccati equation gives, for every $z$ and $v$, the identity $z^\top Qz+rv^2+(Az+Bv)^\top P(Az+Bv)=h(z)+(r+x)\,(v+K^*z)^2$.
Now condition on $\mathcal H^{\mathrm{LQ}}_\tau$ and take expectations of $h(z_{\tau+1})$.
The martingale-difference property \cref{eq:disturbance-assumption} eliminates the cross term in $w_\tau$, and the constant conditional variance makes the noise term equal to $B_w^\top PB_w\,\sigma_w^2=P_{11}\sigma_w^2$ for every policy.
Hence $\mathbb E\bigl[q\epsilon_\tau^2+rv_\tau^2\bigr]+\mathbb E\,h(z_{\tau+1})=\mathbb E\,h(z_\tau)+P_{11}\sigma_w^2+(r+x)\,\mathbb E\bigl[(v_\tau+K^*z_\tau)^2\bigr]$.
Summing over $\tau<T$, telescoping, and dividing by $T$ shows that the running average cost equals $P_{11}\sigma_w^2$ plus a nonnegative excess plus the boundary term $\bigl(\mathbb E\,h(z_0)-\mathbb E\,h(z_T)\bigr)/T$.
Because $P\succeq0$ and the state has finite time-averaged second moments, $\mathbb E\,h(z_T)/T$ vanishes along a subsequence (otherwise those time averages would diverge), and the $\limsup$ in \cref{eq:lqr-cost} dominates the liminf along that subsequence.
Therefore $J(\mu)\ge P_{11}\sigma_w^2$ for every $\mu\in\mathcal M_{\mathrm{LQ}}$.
For the policy \cref{eq:lqr-policy} the excess term vanishes identically and the stable closed loop keeps $\mathbb E\,h(z_\tau)$ bounded, so the limit exists and equals $J^*=P_{11}\sigma_w^2$.
Finally, $\sigma_w^2$ enters neither the Riccati equation nor $K^*$, which is the certainty-equivalence claim.
\end{proof}

\section{Fixed Reporting Delay and Economic-Balance Regulation}
\label{app:delay}

Fix a reporting delay $\delta\ge0$ and known initial $V_0,F_{-1}$, with no costs awaiting reports. Hats denote reported quantities. The economic balance obeys \cref{eq:vault-recurrence}, while
\[
 \hat C_\tau=\begin{cases}0,&\tau<\delta,\\C_{\tau-\delta},&\tau\ge\delta,\end{cases}
 \qquad \hat V_{\tau+1}=\hat V_\tau+F_\tau\bar\Lambda-\hat C_\tau,
 \quad \hat V_0=V_0.
\]
For $\tau\ge\delta$, the unreported liability and reported balance satisfy
\begin{equation}
 \label{eq:inflight}
 L_\tau:=\hat V_\tau-V_\tau=\sum_{s=\tau-\delta}^{\tau-1}C_s,
 \qquad
 \hat V_\tau=V_{\tau-\delta}+\sum_{s=\tau-\delta}^{\tau-1}F_s\bar\Lambda.
\end{equation}
For $\tau<\delta$, the sums start at $s=0$ and the first term is the initial balance $V_0$, since no cost has been reported yet. Thus reports and known fee history reconstruct the economic balance only through $\tau-\delta$. The available information is
\[
 \mathcal H^\delta_\tau:=\sigma(\epsilon_0,\ldots,\epsilon_{\tau-\delta};\,
       \tilde F_{-1},\ldots,\tilde F_{\tau-1}),\qquad \tau\ge\delta,
\]
with only initial error and fee history before $\delta$. Let $\mathcal M^\delta_{\mathrm{LQ}}$ be the policies adapted to this history with the moment conditions of \cref{app:proofs}. There are no additional L1 observations. Assume \cref{eq:disturbance-assumption} relative to the full economic history $\mathcal H^{\mathrm{LQ}}_\tau$.

For $\tau\ge\delta$, define
\begin{equation}
 \label{eq:eps-ce}
 \hat\epsilon_\tau:=\frac{V^*-\hat V_\tau}{V^*},\qquad
 \bar\epsilon_\tau:=\hat\epsilon_\tau+\frac{\delta\bar C}{V^*}
 =\epsilon_{\tau-\delta}-\sum_{s=\tau-\delta}^{\tau-1}u_s.
\end{equation}
During initialization replace $\delta$ by $\tau$ and use $\epsilon_0$ in the last expression. The original normalization $V^*$ and gains remain fixed; only the reported target is raised to provision for unseen costs.

\begin{corollary}[Economic-balance regulation under fixed reporting delay]
\label{cor:delay}
Under these assumptions, the rule of \cref{thm:lqr} applied to $\bar\epsilon_\tau$, with unchanged $\pi,k$, minimizes the economic-balance objective over $\mathcal M^\delta_{\mathrm{LQ}}$. It attains $J^*_\delta=J^*+q\delta\sigma_w^2$, with
\[
 \lim_{T\to\infty}\mathbb E[\mathrm{ISE}_T]=\mathrm{ISE}(\pi)+\delta\sigma_w^2,
 \qquad
 \lim_{T\to\infty}\mathbb E[\mathrm{QV}_T]=\mathrm{QV}(\pi).
\]
The limiting mean economic balance is $V^*$ and the limiting mean reported balance is $V^*+\delta\bar C$.
\end{corollary}

\begin{proof}
By induction on the delayed policy's decisions, its fee history through $\tau-1$ is a function of the fixed initial conditions and economic errors through $\tau-\delta$. Hence $\mathcal H^\delta_\tau\subseteq\mathcal H^{\mathrm{LQ}}_{\tau-\delta}$. Let $e_\tau=\sum_{s=\tau-\delta}^{\tau-1}w_s$. For each summand, the tower property gives conditional mean zero and conditional second moment $\sigma_w^2$ given $\mathcal H^\delta_\tau$. For $s<t$, the dynamics make $w_s$ measurable in $\mathcal H^{\mathrm{LQ}}_t$, so conditioning first on that history gives $\mathbb E[w_sw_t\mid\mathcal H^\delta_\tau]=0$. Therefore
\[
 \epsilon_\tau=\bar\epsilon_\tau+e_\tau,\quad
 \mathbb E[e_\tau\mid\mathcal H^\delta_\tau]=0,\quad
 \mathbb E[e_\tau^2\mid\mathcal H^\delta_\tau]=\delta\sigma_w^2.
\]
Thus $\bar\epsilon_\tau$ is the conditional-mean error and
$\mathbb E[\epsilon_\tau^2]=\mathbb E[\bar\epsilon_\tau^2]+\delta\sigma_w^2$.
Every policy pays an additional $q\delta\sigma_w^2$ in average cost; QV depends only on its applied fees. The finite initialization contributes zero to the long-run average.

The estimate evolves as
$\bar\epsilon_{\tau+1}=\bar\epsilon_\tau-u_\tau+w_{\tau-\delta}$.
The same tower argument gives conditional mean zero and variance $\sigma_w^2$ for its driving disturbance. Given fee history, estimate history and delayed economic-error history determine each other. Their time-averaged moment conditions are equivalent because the estimation error has bounded second moment. The estimate problem is therefore the integrator problem of \cref{thm:lqr}, with the same gains and optimal cost $J^*$. Adding the constant penalty proves the result. Stability gives a limiting mean estimate of zero, and hence the stated balance targets. Only conditional moments, not Gaussianity, are used.
\end{proof}

\subparagraph{If the reported balance itself is the target.}
Applying LQR directly to $\hat\epsilon_\tau$ instead gives limiting mean economic balance $V^*-\delta\bar C$. Its economic tracking error is
\[
 \lim_{T\to\infty}\mathbb E[\mathrm{ISE}_T]
 =\mathrm{ISE}(\pi)+\delta\sigma_w^2+(\delta\bar C/V^*)^2,
\]
with the same limiting QV. These are expected long-run statements; no additive maximum-shortfall formula or identity of realized trajectories follows.

\section{Position Form of the Arbitrum-Style Rule}
\label{app:arbitrum}

The pinned Nitro update allocates a pricing volume $U_{\rm upd}$ to each processed report from its timestamps and accumulated pricing units \cite{nitro_2024_source}. Our adaptation sets $U_{\rm upd}=16\bar\Lambda$ pricing units and converts prices to modeled byte units, giving $\eta=16\bar\Lambda/E$ and $c=\nu\eta/(1+\nu\eta)$. \cref{eq:arb-update} is a real-arithmetic abstraction of the surplus update: integer division, variable allocation, and actual compressed transaction sizes are not reproduced. The historical evaluation applies a zero floor to the updated fee.

Both reference parameters were read from Arbitrum One's pricing getters at block $503{,}764{,}676$ (10 September 2026, 16:22:45 UTC): $E=500{,}000{,}000$ pricing units and $\nu=10$. This verified snapshot supplies one fixed reference; it does not reconstruct parameter changes during the 2025 historical evaluation window.

Without clipping, summing the increments from $s=0$ to $\tau$ telescopes the surplus differences:
\[
 F_\tau=F_{-1}+\frac{c}{\bar U}S_{-1}
       -\frac{c}{\mathrm{EquilUnits}}\sum_{s=0}^{\tau}S_s
       -\frac{c}{\bar U}S_\tau.
\]
The initial fee and initial surplus difference determine the constant term. Substituting $S_\tau=-V^*\epsilon_\tau$ gives proportional gain $K_p^{\mathrm A}=cV^*/\bar U$ and integral gain $K_i^{\mathrm A}=cV^*/\mathrm{EquilUnits}$. With $\bar U=\bar\Lambda$, their normalized values are $c$ and $\eta c$.

\section{Limiting Metrics for PI and LQR}
\label{app:pi-covariance}

\subparagraph{Characteristic polynomials and stability.}
The PI closed-loop matrix below has characteristic polynomial
\begin{equation}
  \label{eq:pi-charpoly}
  \lambda^2-(2-\kappa_p-\kappa_i)\lambda+(1-\kappa_p),
\end{equation}
whose roots lie strictly inside the unit circle exactly when $\kappa_p\in(0,2)$, $\kappa_i>0$, and $2\kappa_p+\kappa_i<4$, giving \cref{eq:pi-stable}.

\subparagraph{PI moments.}
Under the disturbance benchmark \cref{eq:disturbance-assumption}, the PI loop of \cref{sec:ctrl-feedback} has
\begin{align}
  \mathrm{ISE}
  &=\frac{2\sigma_w^2}{\kappa_p\,(4-2\kappa_p-\kappa_i)},
  \label{eq:pi-ise}\\
  \mathrm{QV}
  &=\frac{2\sigma_w^2}{\gamma^2}\,
    \frac{2\kappa_p^3+3\kappa_p^2\kappa_i+\kappa_p\kappa_i^2+\kappa_i^2}
    {\kappa_p\,(4-2\kappa_p-\kappa_i)}.
  \label{eq:pi-qv}
\end{align}

Work in the loop gains $\kappa_p=\gamma K_p$, $\kappa_i=\gamma K_i$ of \cref{sec:ctrl-feedback} and abbreviate the stability margin $D:=4-2\kappa_p-\kappa_i>0$.
The closed loop has state $x_\tau=(\epsilon_\tau,I_{\mathrm{acc},\tau-1})$ and dynamics $x_{\tau+1}=Mx_\tau+e_1w_\tau$ with
\[
  M=\begin{pmatrix}1-\kappa_p-\kappa_i&-\kappa_i\\1&1\end{pmatrix},
  \qquad
  e_1=\begin{pmatrix}1\\0\end{pmatrix}.
\]
Inside the stability triangle \cref{eq:pi-stable}, the state moments converge under the disturbance assumptions \cref{eq:disturbance-assumption}. At the zero-mean second-moment limit, the covariance $S=\mathbb E[x_\tau x_\tau^\top]$ solves the discrete Lyapunov equation $S=MSM^\top+\sigma_w^2\,e_1e_1^\top$, the linear equation that the stationary covariance of any stable linear recursion satisfies.
This is a linear system in the three unknowns $S_{11},S_{12},S_{22}$, with solution
\[
  S_{11}=\frac{2\sigma_w^2}{\kappa_pD},
  \qquad
  S_{12}=-\frac{\sigma_w^2}{\kappa_pD},
  \qquad
  S_{22}=\frac{(2-\kappa_p)\,\sigma_w^2}{\kappa_i\kappa_pD}.
\]
The first entry is \cref{eq:pi-ise}.
The fee increment mixes the current and previous error, $\Delta F_\tau=(K_p+K_i)\epsilon_\tau-K_p\epsilon_{\tau-1}$, so its second moment also needs the lag-one autocovariance of the error.
The lag-one matrix is $\mathbb E[x_{\tau+1}x_\tau^\top]=MS$, whose $(1,1)$ entry gives
\[
  \ell_1:=\mathbb E[\epsilon_\tau\epsilon_{\tau-1}]
  =(1-\kappa_p-\kappa_i)S_{11}-\kappa_iS_{12}
  =\frac{(2-2\kappa_p-\kappa_i)\,\sigma_w^2}{\kappa_pD}.
\]
Assembling the increment moment,
\[
  \mathrm{QV}
  =\mathbb E[(\Delta F_\tau)^2]
  =\bigl[(K_p+K_i)^2+K_p^2\bigr]S_{11}-2(K_p+K_i)K_p\,\ell_1,
\]
and substituting $K_p=\kappa_p/\gamma$, $K_i=\kappa_i/\gamma$ yields \cref{eq:pi-qv} after simplification.
As $\kappa_i\to0$, the error and fee moments approach those of the first-order P loop. $S_{11}$ and $\ell_1$ become the variance and lag-one autocovariance of that autoregression, recovering the proportional formulas. Substituting these limits into the expression for QV recovers
the proportional-controller formulas.

\subparagraph{LQR covariance certificate.}
For $x_\tau=(\epsilon_\tau,u_{\tau-1})^\top$, let $a=(1-\pi)^2/(1+\pi)$ and $D_\pi=1+6\pi+\pi^2$. The closed-loop matrix and limiting covariance are
\[
 M_\pi=\begin{pmatrix}1-a&-\pi\\a&\pi\end{pmatrix},\qquad
 S_\pi=\sigma_w^2\begin{pmatrix}
 \dfrac{(1+\pi)(1+4\pi+\pi^2)}{(1-\pi)D_\pi}&\dfrac{\pi(3+\pi)}{D_\pi}\\[5pt]
 \dfrac{\pi(3+\pi)}{D_\pi}&\dfrac{1-\pi^2}{D_\pi}
 \end{pmatrix}.
\]
Direct substitution verifies $S_\pi=M_\pi S_\pi M_\pi^\top+\sigma_w^2e_1e_1^\top$. Its first entry gives \cref{eq:lqr-ise}; the quadratic form of $(a,\pi-1)/\gamma$ gives \cref{eq:lqr-qv}. The determinant of $M_\pi$ is $\pi$, and its characteristic discriminant is $-4\pi(1-\pi)^2/(1+\pi)^2<0$, so its conjugate poles have modulus $\sqrt\pi$ and imaginary part $\sqrt\pi(1-\pi)/(1+\pi)$. Their angle therefore satisfies $\sin\theta=(1-\pi)/(1+\pi)$. As $\pi\to1$, both $\ln(1/\sqrt\pi)$ and $\theta$ behave like $(1-\pi)/2$, which gives the limiting damping ratio $1/\sqrt2$.

\section{Historical Evaluation Specification and Additional Outcomes}
\label{app:sweep}

\Cref{tab:weights} records the gains and analytical coordinates behind the marked settings.

\subparagraph{Metric boundaries and sweep.}
Tracking uses $V_1,\ldots,V_T$; QV uses $F_1-F_0,\ldots,F_T-F_{T-1}$. The terminal fee $F_T$ is computed but not charged, and $F_0-F_{-1}$ is excluded. Peak fees and zero-fee frequencies use only the charged fees $F_0,\ldots,F_{T-1}$.

The sweep combines 600 logarithmically spaced weights in $[10^{-10},10^{-2}]$ with the exact markers, giving 601 settings per tunable family; \cref{fig:tradeoff} shows their outcomes. It does not cover every stable gain.

% GENERATED by experiments.paper_revision; edit the generator, not this file.
\begin{table}[tbp]
\centering\small
\setlength{\tabcolsep}{3.2pt}
\caption{Parameters and unconstrained benchmark coordinates. P, fixed-ratio PI and LQR minimize the objective at $\rho=\gamma^2q/r$; the Arbitrum-style row is the single fixed reference and has no tuning weight. The gain is $\kappa_p$ for P and PI, $\pi$ for LQR, and $c$ for Arbitrum. PI fixes $\kappa_i=0.1\kappa_p$; Arbitrum has $\kappa_i=\eta c$. Coordinates are $X=\mathrm{ISE}/\sigma_w^2$ and $Y=\gamma^2\mathrm{QV}/\sigma_w^2$.}
\label{tab:weights}
\resizebox{\TableWidth}{!}{%
\begin{tabular}{@{}llrrr@{}}
\toprule
Family & $\rho$ & Gain & $X$ & $Y$ \\
\midrule
P & $10^{-2}$ & 0.131 & 4.085 & 0.01835 \\
PI & $10^{-2}$ & 0.1241 & 4.31 & 0.01969 \\
LQR & $10^{-2}$ & 0.6382305 & 3.425 & 0.01104 \\
\midrule
P & $10^{-6}$ & 0.00629 & 79.75 & $3.968\times10^{-5}$ \\
PI & $10^{-6}$ & 0.005353 & 93.66 & $6.004\times10^{-5}$ \\
LQR & $10^{-6}$ & 0.9562621 & 33.55 & $1.118\times10^{-5}$ \\
\midrule
P & $10^{-10}$ & 0.0002924 & 1710 & $8.55\times10^{-8}$ \\
PI & $10^{-10}$ & $9.781\times10^{-5}$ & 5112 & $5.001\times10^{-7}$ \\
LQR & $10^{-10}$ & 0.9955378 & 335.4 & $1.118\times10^{-8}$ \\
\midrule
Arbitrum ref. & -- & 0.1118 & 4.74 & 0.0135 \\
\bottomrule
\end{tabular}%
}
\end{table}

\end{document}